# Creating Electronic Oscillator-based Ising Machines without External Injection Locking

Jaykumar Vaidya, R S Surya Kanthi, Nikhil Shukla*

Department of Electrical & Computer Engineering, University of Virginia, Charlottesville, Virginia 22904, USA

*Email: ns6pf@virginia.edu

**Abstract.** Coupled electronic oscillators have recently been explored as a compact, integrated circuit- and room temperature operation- compatible hardware platform to design Ising machines. However, such implementations presently require the injection of an externally generated second-harmonic signal to impose the phase bipartition among the oscillators. In this work, we experimentally demonstrate a new electronic autaptic oscillator (EAO) that uses engineered feedback to eliminate the need for the generation and injection of the external second harmonic signal to minimize the Ising Hamiltonian. The feedback in the EAO is engineered to effectively generate the second harmonic signal internally. Using this oscillator design, we show experimentally, that a system of capacitively coupled EAOs exhibits the desired bipartition in the oscillator phases, and subsequently, demonstrate its application in solving the computationally hard Maximum Cut (MaxCut) problem. Our work not only establishes a new oscillator design aligned to the needs of the oscillator Ising machine but also advances the efforts to creating application specific analog computing platforms.

**Introduction**

The Ising model, originally developed for spin glass systems[1], has recently experienced renewed attention owing to its application in accelerating computationally hard problems which are still considered intractable to solve using conventional digital computers. This is motivated by the fact that a large number of such problems[2-6] can be directly mapped to the Ising Hamiltonian: $H = - \sum_{i,j}^{N} J_{ij}\sigma_i\sigma_j$, where $\sigma_i$ is the $i^{th}$ spin ($\pm 1$), $J_{ij}$ is interaction coefficient between spin *i* and spin *j*, and *N* is the total number of spins in the system. For instance, consider the combinatorial optimization-based Maximum Cut (MaxCut) problem – the benchmark problem considered in this work- which entails computing a cut that divides the nodes of the graph in two sets (S1, S2) such that the number of common edges among the two sets is as large as possible. This problem can be mapped to $H$ by considering each node of the graph as a spin. When a node belongs to S1 (S2), then it is assigned a value (spin) +1 (-1), respectively; further, considering only binary weights (relevant to unweighted graphs, considered here), $J_{ij} = -1$, if an edge exists between nodes *i* and *j*; else $J_{ij} = 0$. This ensures that if the two adjacent nodes (i.e., nodes connected by an edge) are in the same set, then $J_{ij}\sigma_i\sigma_j = -1$; if they are in different sets, then $J_{ij}\sigma_i\sigma_j = 1$. Thus, computing the MaxCut of the graph entails finding a configuration of nodes (spins) that minimizes $-\sum_{i,j}^{N} J_{ij}\sigma_i\sigma_j$, which is equivalent to minimizing the Ising Hamiltonian ($H$). Consequently, this has motivated an active effort to realize a physical Ising machine that evolves to minimize its energy (proportional to $H$) and attain the ground state, which in turn, should correspond to the optimal solution of the MaxCut problem. Examples of approaches that have been explored to implement Ising machines include quantum annealing (D-Wave)[7-9], optical parametric oscillator based Coherent Ising machines[10-12], processors based on annealing in- / near-memory[13-15], SRAM based Ising machines that rely on CMOS annealing[16], and coupled electronic oscillators- the focus of the present work. Coupled electronic oscillators have recently been investigated as a promising approach for developing Ising machines since oscillators, in principle, can be made low-power, compact[17-27], compatible with room-temperature operation[28,29], and be manufactured using integrated circuit technology[2,3,30].

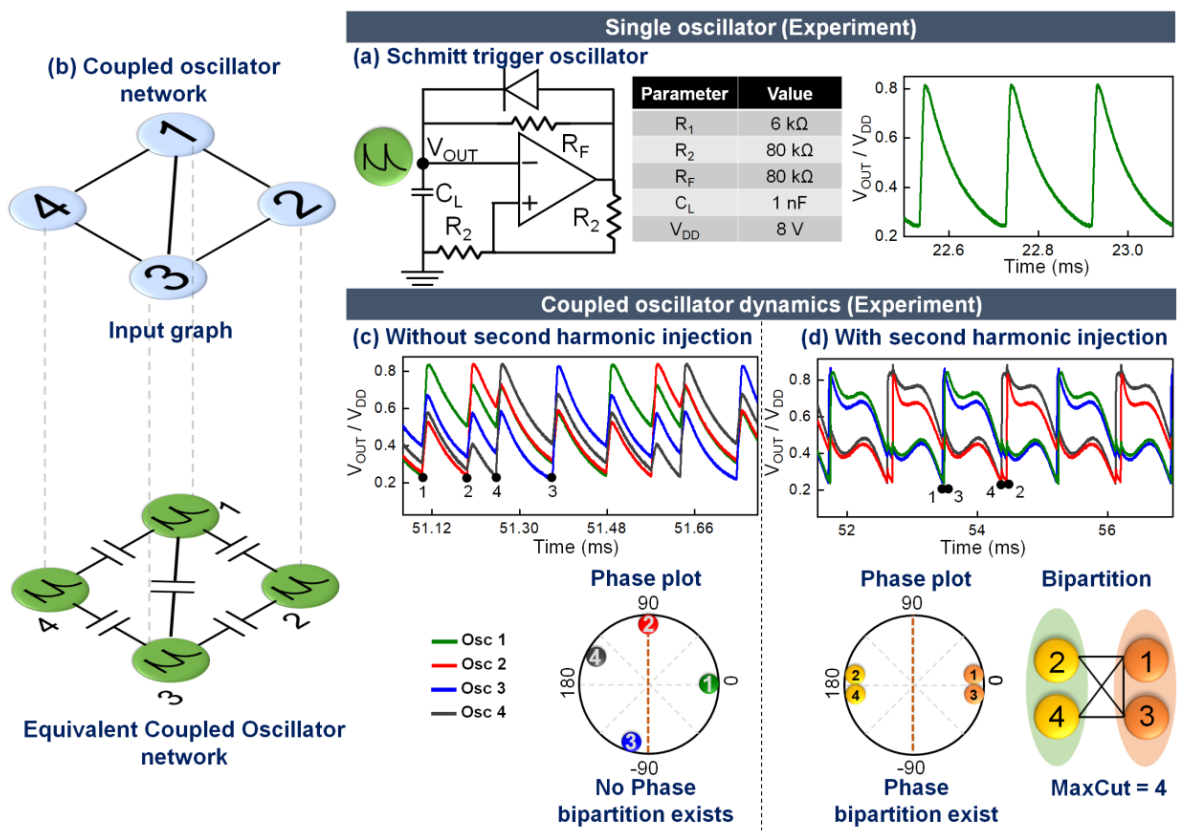

**Fig. 1| Coupled oscillators as an Ising machine.** (a) Schematic of the Schmitt-trigger based relaxation oscillator and the corresponding experimentally measured time domain output. (b) Illustrative graph, and the corresponding coupled oscillator circuit implementation to compute the MaxCut. (c,d) Time domain outputs and corresponding phase plots for a coupled system of oscillators (corresponding to the input graph on the left) with and without external second harmonic injection, respectively. It can be observed that the second harmonic signal is needed to realize the (Ising model-relevant) bipartition in the oscillator phases, required to solve the MaxCut problem. Without the second harmonic signal, the oscillator phases exhibit a multi-partition.

To solve the MaxCut problem directly using the oscillator hardware, each node of the graph is mapped to an oscillator and every edge is represented by a coupling capacitor. The resulting phases exhibit a bipartition (0 or 180º) that can be mapped to the sets S1 and S2 created by the (Max-)Cut. However, coercing the oscillator phases to exhibit the bipartion requires the injection of a second harmonic signal i.e., $f_{inj} \cong 2*f_R$ ($f_{inj}$= frequency of injected signal; $f_R$= frequency of resonant circuit) to every oscillator in the network[4]. In this work, we propose a novel electronic autaptic oscillator (EAO) design that eliminates the need for second harmonic injection; the prefix 'autaptic' is inspired from the autapse structures found in biological spiking neurons which are synapses from the excitatory neuron onto itself (unlike their usual synaptic counterparts that connect to a different neuron) and provide feedback to regulate the neuron's spiking activity[31-34].

**Results**

The role of second harmonic injection is illustrated using the representative 4 node graph shown in Fig. 1. Figure 1a shows the schematic and experimentally measured time domain output of the Schmitt-trigger based relaxation oscillator used in the experiments. The diode in the feedback circuit is used to create an asymmetric output waveform with a rise time that is significantly smaller than the fall time; the motivation for this design feature is explained in the following section. Subsequently, when the oscillators are coupled capacitively ($C_c$= 1nF) in a manner topologically equivalent to the input graph (Fig. 1b), it can be observed that the oscillators are frequency synchronized. Moreover, when no external signal is applied (Fig. 1c), the oscillator phases exhibit a continuous distribution as shown in the corresponding phase plot (the phase ordering, in fact, represents the independent sets of the graph[2,28,29]). However, when an external second harmonic signal ($f_{inj}$=3.4 kHz) is injected (Fig. 1d), the oscillator phases exhibit a bipartition that can subsequently be shown to represent the two sets created by the MaxCut (=4, for the graph considered in the experiment). The second harmonic signal effectively modifies the energy landscape by lowering the energy corresponding to the 0° and 180° oscillator phases[4].

However, generating, and injecting the second harmonic injection signal to every oscillator can incur a significant energy and area overhead. We therefore explore the possibility of eliminating this requirement by redesigning the fundamental compute block of the system- the oscillator. Figure 2a shows the proposed EAO design which essentially consists of a Schmitt-trigger oscillator (similar to that proposed in Fig. 1a) with an additional non-linear feedback element – a diode connected (p-type) MOSFET. Fig. 2b shows the experimentally measured time domain waveform of the oscillator where, unlike the simple oscillator, the discharging phase reveals two distinct two-time constants. While the motivation behind modifying the oscillator dynamics will be discussed in the following sections, we first describe the origin of the two-time constants. The time constant of the oscillator is effectively determined by the net resistance and capacitance in the feedback path. Initially, during the discharging phase, the diode connected p-MOSFET is designed to be in the ON state, and hence, forms a parallel conducting path to the feedback resistance, $R_F$; the diode is reverse biased and has negligible contribution to the conduction. Thus, during this initial phase, the load capacitor, $C_L$, effectively discharges with a time constant $\tau_1 = (R_{FET} || R_F) C_L$. As the output voltage decreases, the diode-connected MOSFET turns OFF, and the output now begins to decay with a larger time constant $\tau_2 = R_F C_L$ ($\tau_2 > \tau_1$) until it reaches the minimum. Subsequently, the oscillator output beings to rise again with the rise time being governed by the dynamic resistance of the diode.

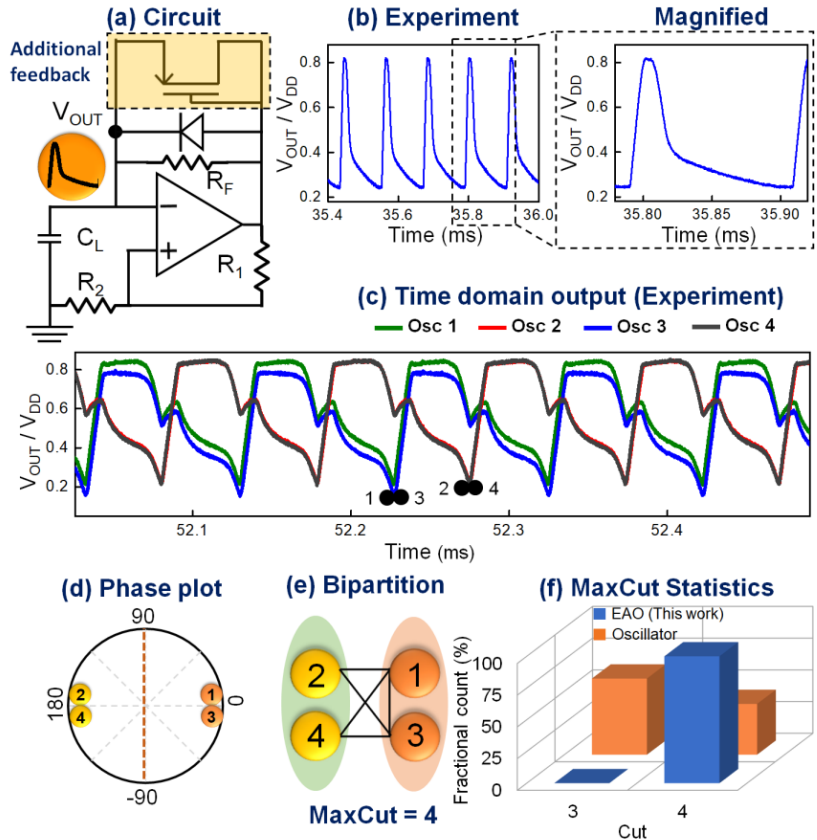

**Fig. 2| Realizing oscillator based Ising machines without second harmonic injection.** (a) Schematic; and (b) time domain output of the proposed EAO. A magnified image of a single oscillation period revealing the two time constants in the relaxation dynamics is also shown. (c) Schematic of the input graph (same as in Fig. 1) and the corresponding experimentally measured time domain output of the equivalent coupled EAO circuit showing the observed phase bipartition. (d) Phase plot; and (e) Corresponding sets (S1,S2) created by the (Max-) Cut; (f) MaxCut solutions measured over 10 runs using network of EAOs, and the conventional oscillators (under second harmonic injection).

Next, we experimentally explore the synchronization dynamics of the EAOs by considering the same representative graph as shown in Fig. 1b, and constructing the corresponding equivalent circuit with the coupled EAOs. It can be observed from Fig. 2e, that the EAO phases exhibit a bipartition (instead of a continuous distribution) without requiring any external injection; in other words, the EAO-based coupled oscillators effectively behave as a system under second harmonic injection, without actually requiring any external injection.

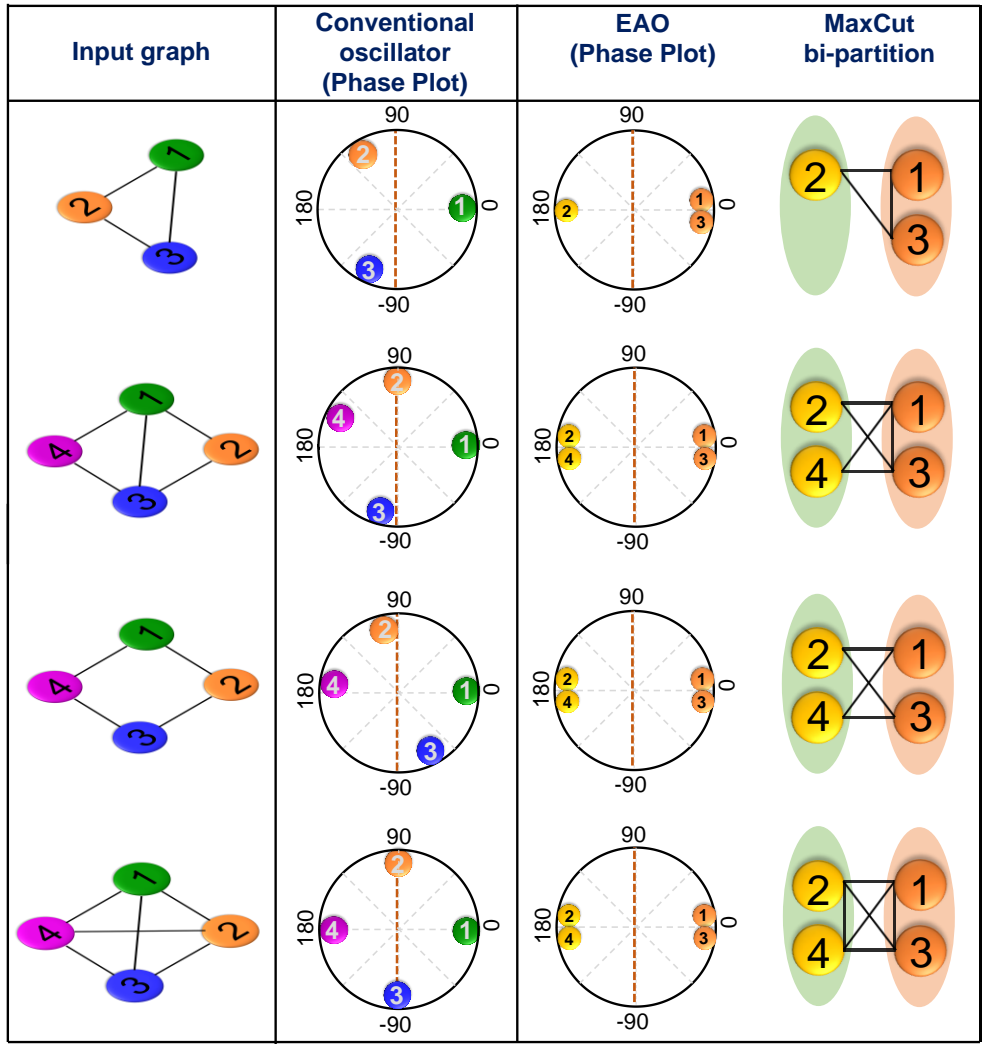

**Fig. 3| Solving MaxCut using coupled EAOs.** Experimentally measured MaxCut solutions for various graph instances obtained using coupled EAOs. The coupled EAOs exhibit a phase bipartition and compute the MaxCut without the need for external second harmonic injection. Corresponding phase dynamics obtained using networks of conventional oscillators (without second harmonic injection) are also shown for comparison. It can be observed that the oscillators do not exhibit a phase bipartition in this case.

Furthermore, the EAOs also compute the optimal solution for the MaxCut of the graph, although there is a statistical distribution (Fig. 2f), similar to that observed with conventional oscillator Ising machines as well. This is not surprising since the system can possibly get trapped in local minima of the high dimensional phase space. Furthermore, we also measure and compare the dynamics of the EAO oscillators with the conventional oscillator design for various graph configurations as shown in Fig. 3. These configurations also exhibit the same bipartition behavior as described earlier as well as show the capability (albeit statistically) to compute the optimal MaxCut solution.

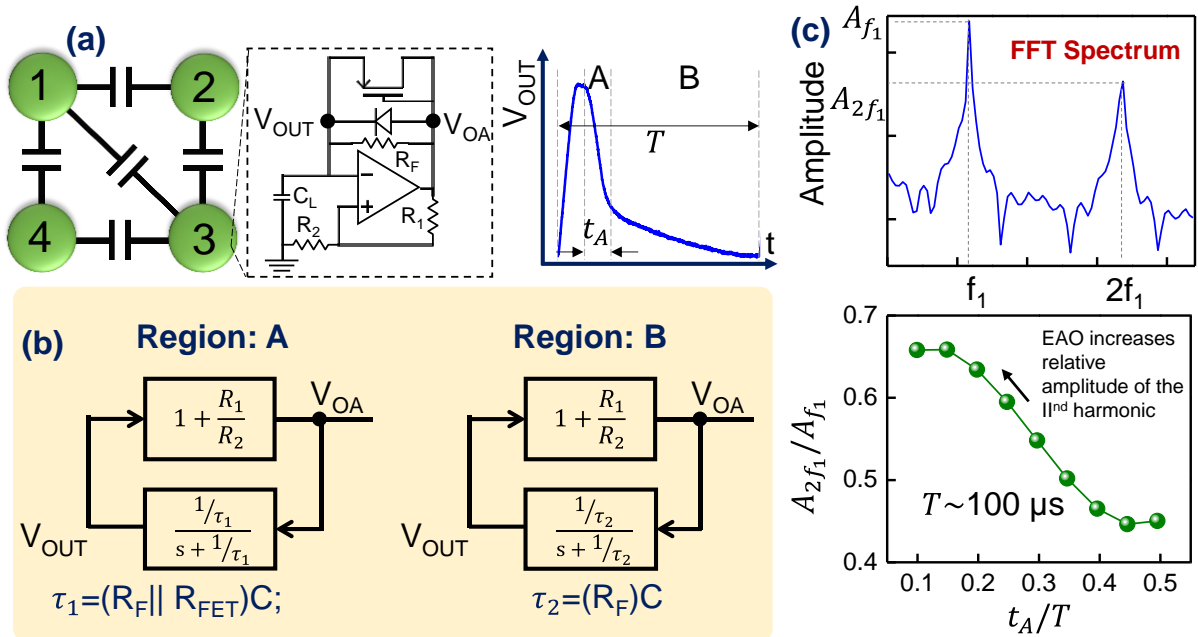

**Fig. 4| Operating principle of the EAO.** (a) Schematic of a coupled EAO circuit with inset showing an EAO. An illustration of the time domain output of the EAO showing the two distinct relaxation regimes (arising from different time constants) is also shown. (b) Control system block diagram of the EAO during the two different phases of voltage relaxation (c) Amplitude of the second harmonic relative to that of the fundamental frequency ($A_{2f_1}/A_{f_1}$) as function of the relative time periods of the two phases observed during voltage relaxation of the EAO.

Next, to analyze the dynamics of the EAO in the coupled circuit, we consider the control block diagram of the EAO (Fig. 4). We specifically focus on voltage relaxation phase in the time domain waveform characterized by two different relaxation time constants. During this phase, the EAO can be modelled as a control system with a feed-forward gain, $G = 1 + \frac{R_1}{R_2}$ and feedback factor $\beta = \frac{1/\tau}{s + 1/\tau}$ where $\tau = RC$. In region A, where the p-MOSFET is considered to be in the ON state, τ₁= (R_F|| R_FET)C_L, and in region B, where the p-MOSFET is OFF, τ₂= R_FC_L (i.e., FET is considered as an open circuit). We note that while the p-MOSFET switching has been considered abrupt between the ON (with ON resistance R_FET) and the OFF state in the above discussion for simplicity, the actual resistance evolution will be a continuum. Using this simplistic model, we analyze the frequency spectrum (Fig. 4c) of the output during the voltage relaxation, specifically focusing on the amplitude of the second harmonic relative to the fundamental frequency, i.e., $A_{2f_1}/A_{f_1}$ (indicative of the relative power that resides at the two frequencies) as a function of the relative time period of the two relaxation phases (expressed as t_A/T; t_A is the time period of the first phase of the voltage relaxation, and T is the total time period of the EAO); the relative time periods are controlled by the RC time constants $\tau_1$ and $\tau_2$. It can be observed from Fig. 4c that $A_{2f_1}/A_{f_1}$ is a strong function of t_A (relative to $T$). In a simple oscillator without the additional feedback, $\tau_1 = \tau_2$ which indicates that the oscillator relaxes with a single time constant. However, as $\tau_2$, and consequently, t_A is progressively reduced using the additional diode-connected transistor element in EAO design, the relative amplitude, and thus, the power concentrated at the second harmonic, steadily increases. We also note that the asymmetric output of the EAO (as well as the conventional oscillator used for comparison), characterized by a short rising time (compared to the voltage relaxation time) helps concentrate more power at the second harmonic in comparison to the symmetric waveform (not shown here).

Thus, by engineering t_A to be small, a significant portion of the EAO power can be concentrated at the second harmonic. Consequently, the EAO effectively self-generates a strong second harmonic signal in the network, and facilitates the bipartition in the oscillator phases without the need for external second harmonic injection.

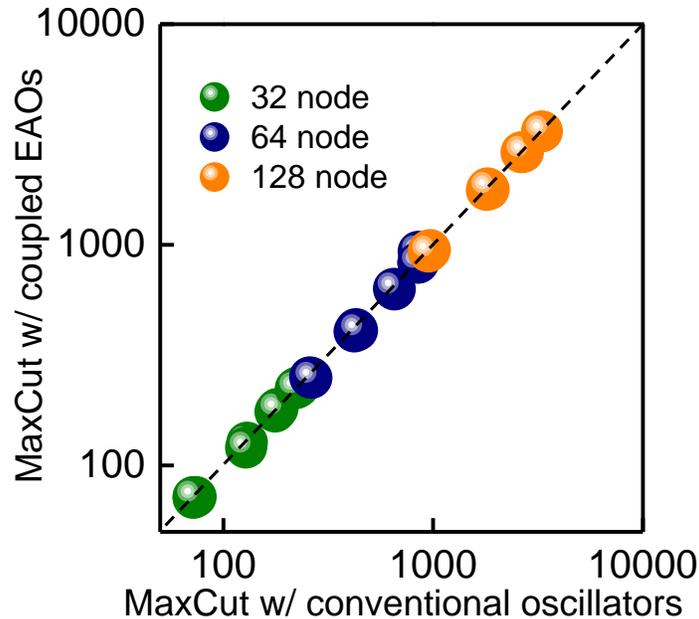

**Fig. 5| Scalability of autaptic oscillator functionality.** Bubble plot comparing the MaxCut solution obtained with the coupled EAOs (without external injection) with that obtained using coupled networks of simple oscillators (under external second harmonic injection). Randomly instantiated graphs of various size (V: 32, 64, 128) and edge density ($\eta$=0.2, 0.4, 0.6, 0.8) are considered; 2 graphs are analyzed for every combination of V, $\eta$; each graph is simulated 10 times. It can be observed that the MaxCut solutions produced by the oscillators are similar to those obtained using simple oscillators.

**Solving larger graphs with EAOs:** Finally, we evaluate using SPICE-based circuit simulations, the functionality of the autaptic oscillators in larger graphs. We analyze randomly generated graph instances of various size (V: 32, 64, 128) and edge densities ($\eta$ = 0.2, 0.4, 0.6, 0.8; $\eta$: ratio of the number of edges in the graph to the number of edges in a complete graph with the same number of nodes) and compare the results with simple oscillators operating under second harmonic injection; 2 graphs are analyzed for every combination of V, $\eta$; each graph is simulated 10 times. Figure 5 shows a bubble plot comparing the MaxCut solution obtained from the coupled EAOs with that obtained using the simple oscillators under the influence of second harmonic injection. It can be observed from the simulations that the autaptic oscillators enable the same computational functionality and similar performance as the simple oscillators under the influence of external second harmonic injection. The EAO based approach exhibits a deviation of -6.2% to +9.15% in comparison to the conventional oscillator-based method ($+^{ive}$ indicates that the EAO solution is better than that produced by the conventional oscillators), with an average deviation of ~-1.6%. This suggests that electronic autaptic oscillator approach can potentially be scaled further although factors such as noise and the coupling architecture will be important to the eventual scalability of the approach.

**Discussion**

The electronic autaptic oscillator concept demonstrated here is a manifestation of a compute-centric device optimization approach to overcome challenges and enable efficient implementation of oscillator-based Ising machines. The feedback methodology used here is not limited to the Schmitt trigger oscillator and can be used to augment other oscillator designs. Considering that area and scalability are key questions in realizing parallel computational architectures like coupled oscillators, this work marks an important step towards improving the scalability and reducing the area requirements for oscillator-based Ising machines.

**References:**

1. Lucas, A. Ising formulations of many NP problems. *Front. Phys.* **2**, 1–14 (2014).

2. Mallick, A., Bashar, M.K., Truesdell, D.S. *et al.* Using synchronized oscillators to compute the maximum independent set. *Nat Commun* **11,** 4689 (2020).

https://doi.org/10.1038/s41467-020-18445-1

3. M. K. Bashar, A. Mallick, D. Truesdell, B. H. Calhoun, S. Joshi, and N. S. Experimental Demonstration of a Reconfigurable Coupled Oscillator Platform to Solve the Max-Cut Problem. 1–6.

4. Wang, T. & Roychowdhury, J. Oscillator-based Ising Machine. (2017).

5. Chou, J., Bramhavar, S., Ghosh, S. & Herzog, W. Analog Coupled Oscillator Based Weighted Ising Machine. *Sci. Rep.* **9**, 1–10 (2019).

6. Albertsson, D.I., Zahedinejad, M., Houshang, A., Khymyn, R., Åkerman, J. and Rusu, A., 2021. Ultrafast Ising Machines using spin torque nano-oscillators. Applied Physics Letters, 118(11), p.112404.

7. M. W. Johnson et al., "Quantum annealing with manufactured spins," Nature, vol. 473, no. 7346, pp. 194–198, May 2011.

8. A. D. King and C. C. McGeoch, "Algorithm engineering for a quantum annealing platform," 2014, arXiv:1410.2628. [Online].

9. V. N. Smelyanskiy et al., "A near-term quantum computing approach for hard computational problems in space exploration," 2012, arXiv:1204.2821. [Online]. Available: http://arxiv.org/abs/1204.2821

10. Wang, Z., Marandi, A., Takata, K., Byer, R. L. & Yamamoto, Y. *A degenerate optical parametric oscillator network for coherent computation. Lecture Notes in Physics* vol. 911 (2016).

11. P. L. Mcmahon et al., "A fully programmable 100-spin coherent ising machine

with all-to-all connections,'' Science, vol. 354, no. 6312, pp. 614–617, Nov. 2016.

12. Y. Haribara, S. Utsunomiya, and Y. Yamamoto, "A coherent ising machine for MAX-CUT problems: Performance evaluation against semidefinite programming and simulated annealing,'' Lect. Notes Phys., vol. 911, pp. 251–262, Oct. 2016.

13. T. Takemoto, M. Hayashi, C. Yoshimura, and M. Yamaoka, "2.6 a 2×30kspin multichip scalable annealing processor based on a Processing-inmemory approach for solving large-scale combinatorial optimization problems,'' in IEEE Int. Solid-State Circuits Conf. (ISSCC) Dig. Tech. Papers, Feb. 2019, pp. 52–54.

14. K. Yamamoto et al., "7.3 STATICA: A 512-spin 0.25 M-weight fulldigital annealing processor with a near-memory all-spin-updates-at-once architecture for combinatorial optimization with complete spin-spin interactions,'' in IEEE Int. Solid-State Circuits Conf. (ISSCC) Dig. Tech. Papers, pp. 138–140, IEEE, 2020.

15. Y. Su, H. Kim, and B. Kim, "31.2 CIM-spin: A 0.5-to-1.2 V scalable annealing processor using digital compute-in-memory spin operators and register-based spins for combinatorial optimization problems,'' in IEEE Int. Solid-State Circuits Conf. (ISSCC) Dig. Tech. Papers, Feb. 2020, pp. 480–482

16. M. Yamaoka, C. Yoshimura, M. Hayashi, T. Okuyama, H. Aoki, and H. Mizuno, "A 20k-spin ising chip to solve combinatorial optimization problems with CMOS annealing,'' IEEE J. Solid-State Circuits, vol. 51, no. 1, pp. 303–309, Jan. 2016.

17. Torrejon, J. et al. Neuromorphic computing with nanoscale spintronic oscillators. Nature 547, 7664 (2017).

18. Lebrun, R. et al. Mutual synchronization of spin torque nano-oscillators through a long-range and tunable electrical coupling scheme. Nat. Commun. 8, 15825 (2017).

19. Coulombe, J. C., York, M. C. & Sylvestre, J. Computing with networks of nonlinear mechanical oscillators. PLoS ONE 12, e0178663 (2017).

20. Csaba, G., Papp, A., Porod, W. & Yeniceri, R. Non-boolean computing based on linear waves and oscillators. In 2015 45th European Solid State Device Research Conference, 101–104 (IEEE, Graz, 2015).

21. Mahboob, I. & Yamaguchi, H. Bit storage and bit flip operations in an electromechanical oscillator. Nat. Nanotechnol. 3, 275–279 (2008).

22. Csaba, G., Ytterdal, T. & Porod, W. Neural network based on parametricallypumped oscillators. In 2016 IEEE International Conference on Electronics, Circuits and Systems (ICECS), 45–48 (IEEE, Monte Carlo, Monaco, 2016).

23. Roychowdhury, J. Boolean computation using self-sustaining nonlinear oscillators. Proc. IEEE 103, 1958–1969 (2015).

24. Pufall, M. R. et al. Physical implementation of coherently coupled oscillator networks. IEEE J. Explor. Solid-State Comput. Devices Circuits 1, 76–84 (2015)

25. G. Csaba and W. Porod, "Noise Immunity of Oscillatory Computing Devices," in IEEE Journal on Exploratory Solid-State Computational Devices and Circuits, vol. 6, no. 2, pp. 164-169, Dec. 2020, doi: 10.1109/JXCDC.2020.3046558.

26. Csaba, G. and Porod, W., 2020. Coupled oscillators for computing: A review and perspective. Applied Physics Reviews, 7(1), p.011302.

27. G. Csaba, A. Raychowdhury, S. Datta and W. Porod, "Computing with Coupled Oscillators: Theory, Devices, and Applications," 2018 IEEE International Symposium on Circuits and Systems (ISCAS), 2018, pp. 1-5, doi: 10.1109/ISCAS.2018.8351664.

28. M. K. Bashar, R. Hrdy, A. Mallick, F. Farnoud Hassanzadeh and N. Shukla, "Solving the Maximum Independent Set Problem using Coupled Relaxation Oscillators," *2019 Device Research Conference (DRC)*, 2019, pp. 187-188, doi: 10.1109/DRC46940.2019.9046422.

29. Vaidya, J., Bashar, M.K. & Shukla, N. Using noise to augment synchronization among oscillators. *Sci Rep* **11,** 4462 (2021). https://doi.org/10.1038/s41598-021-83806-9

30. A. Mallick, M. K. Bashar, D. S. Truesdell, B. H. Calhoun, S. Joshi and N. Shukla, "Graph Coloring Using Coupled Oscillator-Based Dynamical Systems," *2021 IEEE International Symposium on Circuits and Systems (ISCAS)*, 2021, pp. 1-5, doi: 10.1109/ISCAS51556.2021.9401188.

31. Herrmann, C.S. and Klaus, A., 2004. Autapse turns neuron into oscillator. International Journal of Bifurcation and Chaos, 14(02), pp.623-633.

32. Saada, R., Miller, N., Hurwitz, I. and Susswein, A.J., 2009. Autaptic excitation elicits persistent activity and a plateau potential in a neuron of known behavioral function. Current Biology, 19(6), pp.479-484.

33. Yilmaz, E., Ozer, M., Baysal, V. and Perc, M., 2016. Autapse-induced multiple coherence resonance in single neurons and neuronal networks. Scientific Reports, 6(1), pp.1-14.

34. Qin, H., Ma, J., Wang, C. and Chu, R., 2014. Autapse-induced target wave, spiral wave in regular network of neurons. Science China Physics, Mechanics & Astronomy, 57(10), pp.1918-1926.

**Data Availability**

The datasets generated during and/or analyzed during the current study are available from the corresponding author on reasonable request.

**Code Availability**

All codes used in this work are either publicly available or available from the authors upon reasonable request.

**Acknowledgements**

This research was supported in part by the National Science Foundation (Grant No. 1914730).

**Author contributions**

J.V. performed the experiments. J.V. & R.S.S.K. developed and implemented the analytical model and performed the simulations. N.S. conceived the idea and directed the overall project. J.V. and N.S. wrote the manuscript. All authors discussed the results and commented on the manuscript.

**Competing interests**

The authors declare no competing financial interests.